\documentclass[11pt]{article}
\usepackage[T1]{fontenc}
\usepackage{lmodern,microtype}
\usepackage[margin=1in]{geometry}
\usepackage{amsmath,amssymb,amsthm,mathtools}
\usepackage{graphicx,xurl}
\usepackage[section]{placeins}
\usepackage[colorlinks=true,linkcolor=blue,citecolor=blue,urlcolor=blue]{hyperref}
\newtheorem{theorem}{Theorem}[section]
\newtheorem{lemma}[theorem]{Lemma}
\newtheorem{proposition}[theorem]{Proposition}
\theoremstyle{definition}
\newtheorem{remark}[theorem]{Remark}
\title{Payoff-Based Certificates for Convergence\\
to Pure Nash Equilibria under Replicator Dynamics}
\author{Daisuke Hirota}
\date{}
\hypersetup{pdfauthor={Daisuke Hirota},
  pdftitle={Payoff-Based Certificates for Convergence to Pure Nash Equilibria under Replicator Dynamics}}
\begin{document}
\maketitle
\begin{abstract}
For finite bimatrix games, we give a payoff-based sufficient condition $E>0$
for an interior initial state to converge to a specified strict pure Nash
equilibrium under replicator dynamics. The construction combines lower bounds
on the growth rates of log probability ratios between target and alternative
actions with a first integral of a two-dimensional comparison system. It allows
one player's payoff advantage to compensate for the other's disadvantage,
certifying convergence even when a target action initially loses probability.
For rational inputs and integration base points with algebraic exponentials,
we give a finite exact test. We obtain closed-form box probabilities
and, when all nontarget actions are included, closed-form upper bounds on
convergence time. In two-action coordination games, choosing the base points
at the sign thresholds recovers the full basin of attraction. Experiments on
random games show that basin coverage shrinks rapidly with the number of actions.
For $n=4,5,6$, median entry times are approximately 70\% of median box-entry
times without shifts and 40\% for a union over 16 shift choices. Extensions cover
multiplayer games through cooperative comparison systems and explicit boxes,
and discrete-time exponential-weights learning under explicit step-size conditions.
\end{abstract}

\section{Introduction}
Local stability and a basin of attraction answer different questions. Local
stability guarantees convergence to an equilibrium from some neighbourhood
\cite[Theorem~5.2]{MertikopoulosSandholm2016}, but it does not determine whether a
specified initial state converges to that equilibrium. The certificate developed
here supplies such a test. For example, the payoffs determine how close
one player must be to the target to ensure convergence even when the other is far
away. They also allow convergence to a particular equilibrium to be certified
while learning is still in progress.

\paragraph{Contributions.}
\begin{enumerate}
\item \emph{A certificate with compensation.} For any finite bimatrix game and
any strict pure target equilibrium, we construct a payoff-based certificate $E$.
The set where $E>0$, which we call the \emph{capture region}, is an open
inner approximation of the target basin (Theorem~\ref{thm:capture}). It extends beyond the box in which both
players' lower envelopes are positive; Section~\ref{sec:example} gives an example
outside the class of potential games. The test requires neither an enumeration
of Nash equilibria nor global convergence.
\item \emph{Exact computation and closed-form expressions.} For rational data,
membership can be decided by a finite exact computation
(Propositions~\ref{prop:prefix} and~\ref{prop:exacttest}). Box probabilities have
closed-form expressions (Theorem~\ref{thm:volume}), and when the selected sets include every alternative
action, convergence times admit closed-form upper bounds
(Theorem~\ref{thm:time}).
\item \emph{Basin coverage.} In two-action coordination games, threshold
base points give a capture region equal to the basin
(Proposition~\ref{prop:twobytwo}). Every target-convergent initial state enters
the capture region in finite time (Remark~\ref{rem:pullback}). In numerical
experiments, the capture region shrinks rapidly as the number of actions
increases, but for $n=4,5,6$ its median entry time remains approximately 40--70\%
of the median box-entry time (Section~\ref{sec:coverage}).
\item \emph{Extensions.} For multiplayer games, we obtain capture regions
defined by a comparison system and explicit boxes
(Theorem~\ref{thm:multi} and Proposition~\ref{prop:multibox}). The results extend
to discrete-time exponential-weights learning under explicit step-size
conditions (Proposition~\ref{prop:hedge}).
\end{enumerate}

\paragraph{Related work.}
Explicit basin geometry and volume estimates have an extensive literature.
Panageas and Piliouras \cite[Sections~4.3 and~5]{PanageasPiliouras2016} derive
cross-entropy invariants, exact Stag Hunt basins and upper and lower bounds from invariant
polytopes in potential and network coordination games.
Zhang and Hofbauer \cite{ZhangHofbauer2015} study basin dominance, centroid
dominance and risk dominance in two-action bimatrix coordination games.
An invariant of replicator dynamics, expressed as a constant of integration,
implicitly describes the boundary between the two basins, which is the stable
manifold of the saddle. The basin areas are given by integrals
\cite[Eqs.~(10)--(12), (17)]{ZhangHofbauer2015}.
They also note that even centroid dominance should not be expected to have a
simple formula in three-action symmetric coordination games.
For two-action coordination games, the capture region constructed here
coincides with the basin delimited by this boundary
(Proposition~\ref{prop:twobytwo}). The construction can be viewed as extending the
known two-action invariant to games with more actions through a lower comparison
system.
Castro \cite[Theorems~3.1 and~3.4]{Castro2018} shows that, in one-population
matrix games, the basins of replicator and best-response dynamics overlap or
coincide under invariance conditions on sectors bounded by indifference sets.
Dal B\'o and Pujals \cite[Theorem~9]{DalBoPujals2020} give an explicit,
payoff-dependent subset of the basin around a strict equilibrium vertex in
one-population replicator dynamics, defined by an upper bound on the total
probability of alternative actions. Here the same type of guarantee is obtained
for two-population games, with compensation between players enlarging the region.

The contribution is to construct lower envelopes that can be evaluated by a finite computation for each
game and combine them with a first integral of a separable comparison system,
thereby certifying initial states that benefit from compensation.
The multiplayer comparison uses standard order-comparison arguments
\cite[Section~3.1]{HirschSmith2006}.
The cited theorem, section and equation numbers refer to the author versions
listed in the bibliography, including the specified arXiv versions.

\section{Payoff-based convergence certificates}\label{sec:receiver}
This section constructs two one-dimensional integrals from the payoff matrices and uses the sign of their sum $E$ to give
a sufficient condition for an initial state to lie in the basin of the target equilibrium (Theorem~\ref{thm:capture}).

Let $I=\{0,\ldots,n-1\}$ and $J=\{0,\ldots,m-1\}$, with $n,m\ge2$.
Write $\Delta_I=\{p\in\mathbb R_+^I:\sum_i p_i=1\}$ and define
$\Delta_J$ analogously. The domain of initial states is
$D^\circ=\operatorname{int}\Delta_I\times\operatorname{int}\Delta_J$,
where interiors are taken relative to the affine hulls.
Throughout, the measure of initial states is the usual relative Lebesgue measure on $D^\circ$.
For real payoff matrices $A,B\in\mathbb R^{I\times J}$, consider
\begin{equation}\label{eq:replicator}
 \dot p_i=p_i((Aq)_i-p^\top Aq),\qquad
 \dot q_j=q_j((B^\top p)_j-p^\top Bq).
\end{equation}
Time is denoted by $\tau$. The flow $\phi_\tau$ exists at every finite positive and negative time
and preserves $D^\circ$.
Fix a strict pure Nash equilibrium $(t,u)\in I\times J$ and set
\[
 a_k=A_{tu}-A_{ku}>0,\quad c_{kl}=A_{tl}-A_{kl}\quad(k\ne t,l\in J),
\]
\[
 b_l=B_{tu}-B_{tl}>0,\quad d_{lk}=B_{ku}-B_{kl}\quad(l\ne u,k\in I)
\]
so that $c_{ku}=a_k$ and $d_{lt}=b_l$.

\paragraph{Admissible action sets.}
Choose nonempty sets $D_r\subseteq I\setminus\{t\}$ and
$D_c\subseteq J\setminus\{u\}$, and write
$N_r=I\setminus D_r$ and $N_c=J\setminus D_c$. The actions in $D_r,D_c$ will be called \emph{dangerous actions}. We require:
\begin{enumerate}
\item For every $k\in D_r$ and $l\in N_c$, we have $c_{kl}>0$.
\item For every $l\in D_c$ and $k\in N_r$, we have $d_{lk}>0$.
\item In the subgame on $N_r\times N_c$, there exists a finite sequence of eliminations
by pure strict dominance that leaves only $(t,u)$.
\end{enumerate}
In the third condition, each eliminated action is strictly dominated by a pure strategy that is still present;
the dominating strategy has a strictly positive payoff advantage against every remaining opposing action.
The full sets $D_r=I\setminus\{t\}$ and $D_c=J\setminus\{u\}$ are always
admissible. Fix shifts $\kappa\in\mathbb R^{D_r}$ and
$\mu\in\mathbb R^{D_c}$, and define
\[
 \xi_k=\log(p_t/p_k),\quad \eta_l=\log(q_u/q_l),\qquad
 X=\min_{k\in D_r}(\xi_k-\kappa_k),\quad
 Y=\min_{l\in D_c}(\eta_l-\mu_l).
\]

\paragraph{Lower envelopes of payoff differences.}
For $y\in\mathbb R$, let
\[
 R(y)=\{\rho\in(0,\infty)^{J\setminus\{u\}}:
       \rho_l\le e^{-\mu_l-y}\ (l\in D_c)\}
\]
and define
\begin{equation}\label{eq:F}
 F(y)=\min_{k\in D_r}\inf_{\rho\in R(y)}
 \frac{a_k+\sum_{l\ne u}c_{kl}\rho_l}{1+\sum_{l\ne u}\rho_l}.
\end{equation}
Similarly, set
\[
 Q(x)=\{\sigma\in(0,\infty)^{I\setminus\{t\}}:
       \sigma_k\le e^{-\kappa_k-x}\ (k\in D_r)\},
\]
\[
 G(x)=\min_{l\in D_c}\inf_{\sigma\in Q(x)}
 \frac{b_l+\sum_{k\ne t}d_{lk}\sigma_k}{1+\sum_{k\ne t}\sigma_k}.
\]
Throughout, we use the conventions $\min\emptyset=+\infty$ and $\sup\emptyset=-\infty$.

\begin{lemma}\label{lem:envelope}
The envelope has the following finite representation:
\begin{equation}\label{eq:finite-envelope}
\begin{aligned}
 F(y)=\min_{k\in D_r}\min\Bigg\{&
 \min_{T\subseteq D_c}
 \frac{a_k+\sum_{l\in T}c_{kl}e^{-\mu_l-y}}
      {1+\sum_{l\in T}e^{-\mu_l-y}},\\
 &\min_{l\in N_c\setminus\{u\}}c_{kl}\Bigg\}.
\end{aligned}
\end{equation}
It is continuous, locally Lipschitz, and nondecreasing, and
\[
 F(+\infty)=\min_{k\in D_r}
 \min\{a_k,\min_{l\in N_c\setminus\{u\}}c_{kl}\}>0.
\]
The analogous statements hold for $G$ after interchanging the players.
\end{lemma}
\begin{proof}
The fraction in \eqref{eq:F} is a linear-fractional function of $\rho$, so its infimum is given
by vertex values or by limits as unconstrained coordinates tend to infinity.
The limit as an unconstrained coordinate $\rho_l$ ($l\in N_c\setminus\{u\}$) tends to infinity is $c_{kl}$.
The other candidates are obtained by setting the unconstrained coordinates to 0 and each bounded coordinate to either 0 or its upper endpoint.
All these candidates are limits of admissible positive vectors, which proves equality in \eqref{eq:finite-envelope}.
Continuity and local Lipschitz continuity follow from the finite formula. For $y_2\ge y_1$, the inclusion
$R(y_2)\subseteq R(y_1)$ gives monotonicity.
Taking limits in the finite formula gives the stated positive limit by admissibility.
\end{proof}

Define the sign thresholds of the envelopes by
\[
 \bar X_0=\sup\{x:G(x)<0\},\qquad
 \bar Y_0=\sup\{y:F(y)<0\}
\]
and choose finite base points $X_0\ge\bar X_0$ and $Y_0\ge\bar Y_0$.
Such choices exist by Lemma~\ref{lem:envelope}. Define the oriented integrals
\begin{equation}\label{eq:barrier}
 \Gamma(x)=\int_{X_0}^{x}|G(s)|\,ds,\qquad
 \Lambda(y)=\int_{Y_0}^{y}|F(s)|\,ds,\qquad
 E(p,q)=\Gamma(X)+\Lambda(Y),
\end{equation}
and the capture region $\mathcal S=\{(p,q)\in D^\circ:E(p,q)>0\}$.
Smaller admissible finite base points give a larger capture region.
When both thresholds $\bar X_0,\bar Y_0$ are finite, choosing them as the base points maximizes the capture region.

\paragraph{Idea of the construction.}
As shown in the proof below, the growth rates of $X,Y$ are bounded below by $F(Y)$ and $G(X)$, respectively (equation~\eqref{eq:comparison}).
Replacing the inequalities with equalities gives the comparison system $\dot x=F(y)$, $\dot y=G(x)$.
This system is separable and preserves $\int G\,dx-\int F\,dy$.
The function $E$ aligns the signs in this first integral in favor of the target. Indeed, in a region where exactly one envelope is negative
(for example, $G(x)<0<F(y)$), the comparison system satisfies
$\dot E=|G(x)|F(y)+|F(y)|G(x)=0$.
Thus, even if one coordinate lies below its threshold, $E>0$ can hold when the other integral is sufficiently large.
This is the compensation between the two players.

\begin{theorem}\label{thm:capture}
For any admissible pair of action sets, shifts, and finite base points as above,
$\mathcal S$ is a nonempty open forward-invariant subset of $D^\circ$.
Along its trajectories, $E$ is nondecreasing and
$p_t(\tau)\to1$, $q_u(\tau)\to1$.
In particular, $\mathcal S$ has positive relative Lebesgue measure and
is contained in the basin of attraction of $(t,u)$ under \eqref{eq:replicator}.
\end{theorem}
\begin{proof}
Taking differences of logarithmic derivatives gives
\[
 \dot\xi_k=(Aq)_t-(Aq)_k,
 \qquad \dot\eta_l=(B^\top p)_u-(B^\top p)_l.
\]
Substituting $\rho_l=q_l/q_u$ gives the fraction in \eqref{eq:F}.
Since $\rho\in R(Y)$, every log ratio associated with a dangerous row action has derivative at least
$F(Y)$. Similarly, every log ratio associated with a dangerous column action has derivative at least $G(X)$.
These derivatives are bounded by the finite payoff differences. Hence $X,Y$
are Lipschitz functions of time. The right derivative of the minimum of finitely many $C^1$ functions
equals the smallest derivative among the active indices. Therefore,
\begin{equation}\label{eq:comparison}
 D_+X\ge F(Y),\qquad D_+Y\ge G(X),
\end{equation}
where $D_+$ denotes the lower right Dini derivative.
In particular, these inequalities hold for the ordinary derivatives almost everywhere.

The functions $\Gamma,\Lambda$ are $C^1$ and nondecreasing.
The base point conditions and monotonicity imply
\[
 F(y)<0\Longrightarrow\Lambda(y)<0,\qquad
 \Lambda(y)>0\Longrightarrow F(y)>0,
\]
together with the corresponding implications for $G,\Gamma$.
Moreover, $\Lambda(y)\to\infty$, $\Gamma(x)\to\infty$ as their arguments tend to infinity.
Along any trajectory, $E$ is locally absolutely continuous, and the chain rule gives
\[
 \dot E\ge |G(X)|F(Y)+|F(Y)|G(X)
\]
almost everywhere.
The right-hand side can be negative only when both $F(Y)$ and $G(X)$ are negative.
In that case, both oriented integrals are negative, so $E<0$.
Thus, on every interval on which $E\ge0$, we have $\dot E\ge0$ almost everywhere.
If $E(0)=e_0>0$, integration up to the first possible exit from $\{E>0\}$ shows that
$E(\tau)\ge e_0$ up to that time. Continuity rules out an exit in finite time.
This proves forward invariance and monotonicity on $\mathcal S$.

We next show that both minimum log ratios tend to infinity.
Set
\[
 y_* =\min\{y:\Lambda(y)\ge e_0\},\quad
 x_* =\min\{x:\Gamma(x)\ge e_0\},\quad
 \delta=\min\{F(y_*),G(x_*)\}>0.
\]
These minima exist because the integrals are nonpositive below their finite base points
and tend to infinity above them. If $X<X_0$, then $\Gamma(X)\le0$,
so $\Lambda(Y)\ge e_0$ and $D_+X\ge\delta$.
Similarly, whenever $Y<Y_0$, we have $D_+Y\ge\delta$.
If a coordinate were to fall below its base point after reaching it,
it would have to increase throughout that excursion, a contradiction.
Thus each base point is reached in finite time and is never crossed downward thereafter.
At some finite time $T_0$, both $X\ge X_0$ and $Y\ge Y_0$ hold.
Both coordinates are subsequently nondecreasing by \eqref{eq:comparison}.
Since $E(T_0)>0$, at least one of the integrals is positive. For example,
if $\Lambda(Y(T_0))>0$, then
$D_+X\ge F(Y(T_0))>0$ thereafter. Hence $X\to\infty$, and eventually
$D_+Y\ge G(+\infty)/2>0$. The other case follows by interchanging the players.
Thus, after a finite time, both coordinates grow at least linearly.
The bounds $p_k\le e^{-\kappa_k-X}$ and $q_l\le e^{-\mu_l-Y}$
then imply that the probabilities of all dangerous actions decay exponentially.

Follow a finite elimination sequence witnessing admissibility of the remaining subgame.
Inductively, suppose that, for some $C,\gamma>0$, the total probability of the actions already eliminated, including the dangerous actions,
is bounded by $Ce^{-\gamma\tau}$.
When a row action $k$ is next eliminated because it is dominated by a remaining action $j$,
let the minimum of $A_{jl}-A_{kl}$ over the remaining columns be $h>0$,
and set $M=2\max_{i,l}|A_{il}|$.
Then
\[
 \frac{d}{d\tau}\log\frac{p_j}{p_k}
 \ge h-(h+M)Ce^{-\gamma\tau}\ge h/2
\]
for all sufficiently large $\tau$. Since $p_j\le1$, this yields
an exponential upper bound on $p_k$. The column step is identical.
Adjusting the constants at each of the finitely many steps proves
the decay of all nontarget actions and hence convergence to $(t,u)$.
Finally, $E$ is continuous on the original domain. Taking $p_t,q_u$
sufficiently close to 1 makes both minima and both integrals positive.
Thus $\mathcal S$ is nonempty and open and has positive relative measure.
\end{proof}

\begin{remark}[Order relative to the comparison system]\label{rem:order}
By Lemma~\ref{lem:envelope}, $F,G$ are locally Lipschitz and nondecreasing, so the comparison system
$\dot x=F(y)$, $\dot y=G(x)$ is cooperative: the growth rate of each component is nondecreasing in the other component.
For the solution of the comparison system starting at
$(x(0),y(0))=(X(0),Y(0))$, we have
$X(\tau)\ge x(\tau)$, $Y(\tau)\ge y(\tau)$ for all $\tau\ge0$.
Indeed, lower each component of the comparison vector field by $\varepsilon>0$, and let $(x_\varepsilon,y_\varepsilon)$ solve the resulting system.
By \eqref{eq:comparison} and the monotonicity of $F,G$, the right Dini derivatives of $X-x_\varepsilon$ and $Y-y_\varepsilon$ are at least $\varepsilon$
as long as both differences are nonnegative. A continuous function with positive right Dini derivative is increasing,
so both differences remain positive at positive times. Letting $\varepsilon\to0$ gives the claim.
This is the same argument used to prove order preservation for cooperative systems \cite[Theorem~3.2]{HirschSmith2006}.
The proof of Theorem~\ref{thm:capture} uses only the monotonicity of $E$, without invoking this order relation.
\end{remark}

\section{Example: convergence with compensation in rock--paper--scissors with coordination}
\label{sec:example}
This section gives an example in which the compensation in Theorem~\ref{thm:capture} is effective.
The certificate establishes convergence from an initial state at which the probability of the target action initially decreases.

Add a coordination action $0$ to a three-action constant-sum rock--paper--scissors game. We refer to actions $1,2,3$ as the rock--paper--scissors actions.
When both players choose coordination, each receives $2$;
when one chooses coordination and the other does not, each receives $0$. The payoff matrices are
\begin{equation}\label{eq:example-matrices}
A=\begin{pmatrix}
2&0&0&0\\
0&1&0&2\\
0&2&1&0\\
0&0&2&1
\end{pmatrix},\qquad
B=\begin{pmatrix}
2&0&0&0\\
0&1&2&0\\
0&0&1&2\\
0&2&0&1
\end{pmatrix}.
\end{equation}
Equivalently, the rock--paper--scissors blocks are $\mathbf 1+K$ and
$\mathbf 1-K$, where
$K=\left(\begin{smallmatrix}0&-1&1\\1&0&-1\\-1&1&0\end{smallmatrix}\right)$.
The target $(0,0)$ is a strict pure Nash equilibrium, with unilateral payoff gaps equal to $2$ for both players.
Take $D_r=D_c=\{1,2,3\}$ and set all shifts to 0. These sets are admissible
because their complements contain only the target action.

\begin{figure}[!htbp]
\centering
\includegraphics[width=.72\linewidth]{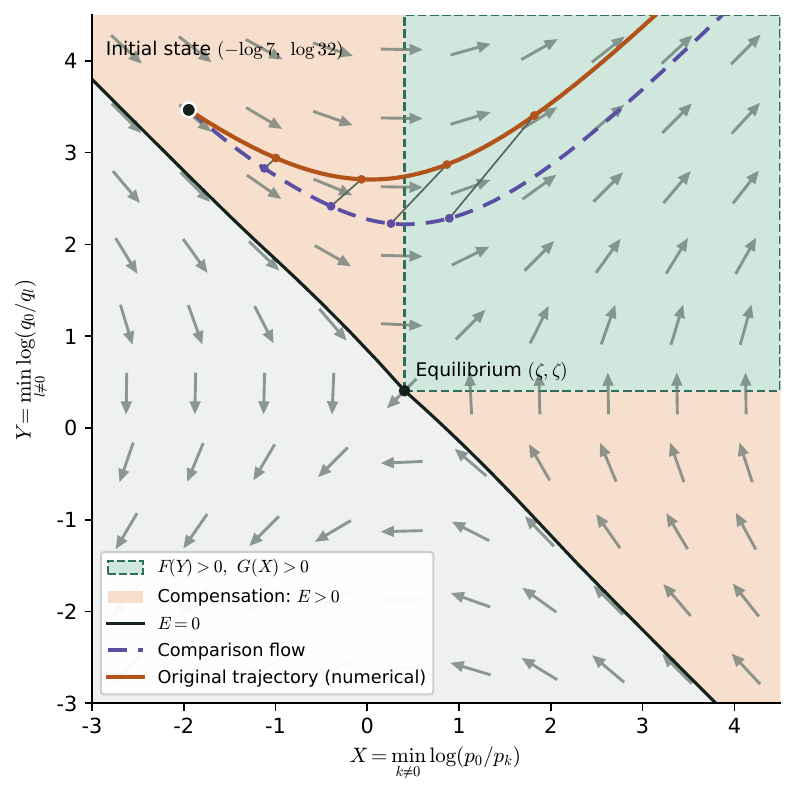}
\caption{Capture region and comparison system for the example in Section~\ref{sec:example}, in minimum log-ratio coordinates. The green shading and dashed lines show
the invariant box $\{F(Y)>0,\ G(X)>0\}$; the orange shading shows the region where compensation gives $E>0$; the black curve is $E=0$;
and the gray arrows show the vector field of the comparison system. The purple dashed curve is the comparison trajectory from the marked initial state, and the orange solid curve is
the original trajectory from the same initial state mapped to $(X,Y)$. Thin lines connect the positions at the same times $\tau=0.5,1,1.5,2$.
Both trajectories are obtained by numerical integration.}
\label{fig:receiver}
\end{figure}

For every dangerous action, the payoff differences between the target and that action against
the opponent's rock--paper--scissors actions are a permutation of $(-2,-1,0)$.
Both envelopes therefore equal $f$, where, writing $s=e^{-y}$,
\begin{equation}\label{eq:example-envelope}
f(y)=\begin{cases}
\dfrac{2-2s}{1+s},&s\ge3,\\[3pt]
\dfrac{2-3s}{1+2s},&2/3\le s\le3,\\[3pt]
\dfrac{2-3s}{1+3s},&0<s\le2/3.
\end{cases}.
\end{equation}
Its unique zero is $\zeta=\log(3/2)$. Use this minimal base point
in both integrals in \eqref{eq:barrier}. This gives
$E=H(X)+H(Y)$, with $H(z)=\int_\zeta^z|f(v)|\,dv$.
Figure~\ref{fig:receiver} shows the capture region and the comparison system. As explained in Remark~\ref{rem:order}, the original trajectory never lies below the comparison trajectory in either coordinate.

\begin{proposition}\label{prop:example}
For the matrices in \eqref{eq:example-matrices}, the interior initial state
\begin{equation}\label{eq:example-source}
p=\frac{(2,1,3,14)}{20},\qquad
q=\frac{(512,16,2,1)}{531}
\end{equation}
converges to $(0,0)$. At this initial state, the probability of the column player's target action
initially decreases, and action $0$ is not a best response for the column player.
These conclusions hold throughout an open neighbourhood of this initial state
in the original six-dimensional positive state space.
\end{proposition}

\begin{proof}
In the finite representation of Lemma~\ref{lem:envelope}, all weights are equal, so the denominator depends only on $|T|$.
For each $|T|$, it therefore suffices to choose the smallest payoff differences (a special case of Proposition~\ref{prop:prefix}).
Thus the envelope reduces to the minimum of
\[
g_0=2,\quad g_1=\frac{2-2s}{1+s},\quad
g_2=\frac{2-3s}{1+2s},\quad g_3=\frac{2-3s}{1+3s}.
\]
After multiplication by the positive denominators, the numerators of $g_1-g_2$, $g_2-g_3$, and $g_1-g_3$ are
$s(3-s)$, $s(2-3s)$, and $s(5-3s)$, respectively.
This proves \eqref{eq:example-envelope}.
At \eqref{eq:example-source}, the active minimizers are unique, and
\[
X=\log(p_0/p_3)=-\log7<\zeta<\log32=\log(q_0/q_1)=Y.
\]
To evaluate $E$ exactly, set $\beta=-\log3$ at the branch boundary and define the primitives
\begin{align*}
J_1(y)&=2y+4\log(1+e^{-y}),\\
J_2(y)&=2y+\tfrac72\log(1+2e^{-y}),\\
J_3(y)&=2y+3\log(1+3e^{-y}).
\end{align*}
Each derivative is the corresponding branch of $f$. Since $f$ is negative below $\zeta$
and positive above $\zeta$,
\[
E=J_1(\beta)-J_1(X)+J_2(\zeta)-J_2(\beta)+J_3(Y)-J_3(\zeta).
\]
Substituting $e^{-X}=7$, $e^{-\beta}=3$, $e^{-\zeta}=2/3$, and $e^{-Y}=1/32$
gives the following exact comparison, an instance of the procedure in Proposition~\ref{prop:exacttest}:
\[
e^{2E}=\frac{5^6 7^{10}}{2^{18}3^{13}}
=\frac{4413675765625}{417942208512}>1.
\]
Hence $E>0$, and Theorem~\ref{thm:capture} gives convergence to the target.

At this initial state, the row fitnesses are $(1024/531,2/59,34/531,5/531)$,
and the column fitnesses are $(1/5,29/20,1/4,1)$. Therefore,
\begin{align*}
\dot X&=\frac{1019}{531}>F(Y)=\frac{61}{35},
&\dot X-F(Y)&=\frac{3274}{18585},\\
\dot Y&=-\frac54>G(X)=-\frac32,
&\dot Y-G(X)&=\frac14.
\end{align*}
In particular, for the column player, action $0$ has lower fitness than every rock--paper--scissors action, and
\[
\dot q_0=-\frac{53504}{1409805}<0.
\]
The inequalities $E>0$ and $\dot q_0<0$, together with the strict payoff inequalities excluding the target action as a best response,
persist on an open neighbourhood by continuity.
\end{proof}

\paragraph{Comparison with other regions.}
This initial state lies outside the box $\{X>\zeta,Y>\zeta\}$ where both envelopes are positive.
It also lies outside the elementary rectangle $\{p_0>1/2,q_0>1/2\}$ in the two target-action probabilities (here $p_0=1/10$).
Against any mixed strategy of the opponent, the payoff differences between the target and a rock--paper--scissors action are at least
$4q_0-2$ and $4p_0-2$, respectively, so this rectangle is
a forward-invariant subset of the basin.
The certificate predicts that $q_0$, which initially decreases, eventually reverses direction.

\paragraph{Absence of a potential.}
This game lies outside the class of potential games \cite[Section~2, Lemma~2.3]{MondererShapley1996}.
Among the rock--paper--scissors action profiles,
\[
(2,1)\to(2,3)\to(1,3)\to(1,2)\to(3,2)\to(3,1)\to(2,1)
\]
is a cycle of unilateral strict improvements, with the deviating player receiving payoff $2$ at every step.
An ordinal potential that strictly increases along this cycle cannot return to its initial value.
Hence no ordinal potential exists, and therefore no exact potential exists.

\paragraph{Convergence to the target is not global.}
At the strategy profile in which both players mix uniformly over the rock--paper--scissors actions,
$e=(0,1/3,1/3,1/3)$, the fitness of each rock--paper--scissors action is $1$,
whereas that of the absent target action is $0$. In fact, the curve
\[
p=q=\left(\alpha,\frac{1-\alpha}{3},
\frac{1-\alpha}{3},\frac{1-\alpha}{3}\right),\qquad 0<\alpha<1/3,
\]
is invariant. Along this curve, $\dot\alpha=\alpha(1-\alpha)(3\alpha-1)<0$,
so every trajectory on it converges to $e$. Thus not every interior initial state converges to the target.

\section{Explicit boxes, volumes, and time bounds}\label{sec:volume}
We now construct boxes inside the capture region, derive closed-form expressions for their probabilities, and obtain upper bounds on convergence times.
These boxes give coarser certificates than the capture region, but all quantities can be computed directly from the payoffs.

For a real number $z$, write $z^-=\max\{-z,0\}$ and $[z]_+=\max\{z,0\}$.
The thresholds in Section~\ref{sec:receiver} have the closed forms
\begin{equation}\label{eq:thresholds}
 \bar X_0=\log\max_{l\in D_c}
 \frac{\sum_{k\in D_r}e^{-\kappa_k}d_{lk}^-}{b_l},\qquad
 \bar Y_0=\log\max_{k\in D_r}
 \frac{\sum_{l\in D_c}e^{-\mu_l}c_{kl}^-}{a_k},
\end{equation}
where $\log0=-\infty$.
Indeed, minimizing each numerator over its bounded coordinates selects exactly the negative coefficients.
All such numerators are positive above the largest root, whereas at least one is negative below it.
The constants associated with unconstrained coordinates are positive.
Thus $G(x)>0$ for $x>\bar X_0$ and $G(x)<0$ for $x<\bar X_0$; the analogous statements hold for $F$.

Let $\nu_D$ be the product of the uniform probability measures on the two simplices.
This is the normalized relative Lebesgue measure on the ambient product space, rather than volume on a lower-dimensional invariant face.
For finite $\bar x>\bar X_0$ and $\bar y>\bar Y_0$, define the following subset of $D^\circ$, which is a box in the selected log-ratio coordinates:
\[
 \mathcal B(\bar x,\bar y)=\{\xi_k\ge\kappa_k+\bar x\ (k\in D_r),\quad
                              \eta_l\ge\mu_l+\bar y\ (l\in D_c)\}.
\]
\begin{theorem}\label{thm:volume}
Every point in $\mathcal B(\bar x,\bar y)$ converges to $(t,u)$, and
\[
 \nu_D(\mathcal B(\bar x,\bar y))=V_r(\bar x)V_c(\bar y),\qquad
 V_r(x)=\sum_{T\subseteq D_r}
 \frac{(-1)^{|T|}}{1+\sum_{k\in T}e^{-\kappa_k-x}},
\]
where $V_c$ is defined by replacing $(D_r,\kappa,x)$ with $(D_c,\mu,y)$.
Consequently, the probability of the target basin is at least
$V_r(\bar X_0)V_c(\bar Y_0)$, where $V_r(-\infty)=V_c(-\infty)=1$
is interpreted as a limit.
Moreover, if $\bar x\ge X_0$ and $\bar y\ge Y_0$, with at least one strict inequality,
then $\mathcal B(\bar x,\bar y)\subseteq\mathcal S$
(within the range $\bar x>\bar X_0$, $\bar y>\bar Y_0$ in which the box is defined).
\end{theorem}
\begin{proof}
As long as $X\ge\bar x$ and $Y\ge\bar y$, their derivatives are bounded below by
$F(\bar y)>0$ and $G(\bar x)>0$, respectively.
The box is therefore invariant, both minima grow linearly, and convergence follows from the elimination argument in the proof of Theorem~\ref{thm:capture}.
For the probability formula, represent a uniformly distributed point on a simplex by independent unit-rate exponential random variables $Z_i$ normalized by their sum.
Writing $h_k=e^{\kappa_k+x}$ and conditioning on $Z_t$ gives
\[
 \Pr(Z_k\le Z_t/h_k\ \forall k\in D_r)
 =\mathbb E\prod_{k\in D_r}(1-e^{-Z_t/h_k})=V_r(x).
\]
Independence across players gives the product.
Lowering the bounds $\bar x,\bar y$ to the thresholds yields the basin bound by continuity from below.
Under the additional base point conditions, both integrals are nonnegative and one is positive.
\end{proof}

\begin{theorem}\label{thm:time}
Let $D_r=I\setminus\{t\}$ and $D_c=J\setminus\{u\}$, and suppose that the initial point lies in
$\mathcal B(\bar x,\bar y)$.
Write $C_r=\sum_{k\ne t}e^{-\kappa_k}$ and $C_c=\sum_{l\ne u}e^{-\mu_l}$.
For every $\tau\ge0$,
\[
 1-p_t(\tau)\le C_r e^{-\bar x-F(\bar y)\tau},\qquad
 1-q_u(\tau)\le C_c e^{-\bar y-G(\bar x)\tau}.
\]
For $0<\varepsilon<1$, the earliest time after which both errors remain at most
$\varepsilon$ is bounded above by
\[
 \max\left\{\frac{[\log(C_r/\varepsilon)-\bar x]_+}{F(\bar y)},\quad
             \frac{[\log(C_c/\varepsilon)-\bar y]_+}{G(\bar x)}\right\}.
\]
\end{theorem}
\begin{proof}
Invariance of the box gives $X(\tau)\ge\bar x+F(\bar y)\tau$ and
$Y(\tau)\ge\bar y+G(\bar x)\tau$.
Sum the individual ratio bounds over all nontarget actions and solve the exponential inequalities.
\end{proof}

\section{Finite evaluation and exact membership tests}\label{sec:computation}
We show that the envelopes $F,G$ can be evaluated using a number of candidates proportional to the number of actions, rather than the number of action subsets
(Proposition~\ref{prop:prefix}), and that, for rational data, the sign of $E$ can be determined by a finite exact computation
(Proposition~\ref{prop:exacttest}).

Fix a row $k\in D_r$, abbreviate $a=a_k$ and $c_l=c_{kl}$, and set $s=e^{-y}>0$ and $w_l=e^{-\mu_l}>0$.
Write
\[
 R_T(s)=\frac{a+sC_T}{1+sD_T},\qquad
 C_T=\sum_{l\in T}w_lc_l,\qquad D_T=\sum_{l\in T}w_l.
\]
The remaining candidates in the envelope are the constants associated with the unconstrained coordinates.

\begin{proposition}[Prefix representation]\label{prop:prefix}
Sort the coefficients as $c_{(1)}\le\cdots\le c_{(d)}$, breaking ties in any fixed order.
Let $C_j,D_j$ be the corresponding prefix sums, including $C_0=D_0=0$.
Then, for every $s>0$,
\[
 \min_{T\subseteq\{1,\ldots,d\}}R_T(s)
 =\min_{0\le j\le d}\frac{a+sC_j}{1+sD_j}.
\]
This order depends on neither $s$ nor the positive weights.
\end{proposition}
\begin{proof}
Choose a minimizing subset $T$ and write $r=R_T(s)$.
For $l\notin T$ and $l\in T$, respectively,
\[
 R_{T\cup\{l\}}(s)-r
 =\frac{sw_l(c_l-r)}{1+s(D_T+w_l)},\qquad
 R_{T\setminus\{l\}}(s)-r
 =\frac{sw_l(r-c_l)}{1+s(D_T-w_l)}.
\]
Both denominators are positive.
Optimality therefore requires that $T$ contain every $l$ with $c_l<r$ and that $T$ contain no $l$ with $c_l>r$.
Removing terms with $c_l=r$ leaves the value unchanged.
The remaining set is a sorted prefix, possibly empty.
\end{proof}

For $r$ rows and $d$ dangerous actions of the opponent, preprocessing sorts each row and computes its prefix sums.
This requires $O(rd\log d)$ comparisons and arithmetic operations when $d\ge2$; the cases $d\le1$ are handled directly.
Reducing the unconstrained-coordinate constants to one minimum per row requires one additional pass through those entries.
Evaluation at a fixed $y$ then uses the unconstrained-coordinate constants and $r(d+1)$ fractional candidates.
A constant arising from an unconstrained coordinate remains a candidate for the infimum even if no positive vector attains it.
The same procedure evaluates $G$.

\begin{proposition}[Exact sign determination]\label{prop:exacttest}
Assume that all payoff entries are rational, the weights $e^{-\mu_l}$ and $e^{-\kappa_k}$ are positive rational numbers,
and all coordinates of the initial point are positive rational numbers.
Suppose further that the given admissible base points satisfy
$e^{-X_0},e^{-Y_0}\in\overline{\mathbb Q}_{>0}$,
with these exponentials specified by defining polynomials and rational isolating intervals.
Then the sign of the quantity $E$ in~\eqref{eq:barrier}, including the case $E=0$,
can be determined by finitely many exact real algebraic operations.
\end{proposition}
\begin{proof}
Each prefix branch or constant branch, expressed in the variable $s=e^{-y}$, has the form
\[
 g(s)=\frac{a+Cs}{1+Ds},\qquad a,C,D\in\mathbb Q,\quad D\ge0.
\]
Here $D=0$ implies $C=0$: a prefix with positive weights must then be empty, and an unconstrained-coordinate branch is constant.
The equation determining intersections of two branches is
\[
 (a_1+C_1s)(1+D_2s)-(a_2+C_2s)(1+D_1s)=0.
\]
Its left-hand side is a polynomial of degree at most two.
Identical branches are merged, and linear and constant cases are handled according to their actual degree.
Also include the positive roots of $a+Cs=0$ for nonzero numerators.
These finitely many algebraic breakpoints partition any compact positive $s$ interval into cells on which the minimizing branch and its sign can be selected exactly by sampling rational interior points.
Ties do not change the value, and an identically zero branch contributes 0.

For $D>0$, an antiderivative with respect to $y$ is
\[
 J(y)=ay+(a-C/D)\log(1+De^{-y}),\qquad
 J'(y)=\frac{a+Ce^{-y}}{1+De^{-y}}.
\]
For $D=0$, an antiderivative is $ay$.
Integrate $|F|$ and $|G|$ using the sign of the branch on each cell, preserving the orientation when the endpoint coordinate lies below its base point.
The transformed endpoints corresponding to the initial point are rational; for example,
\[
 e^{-X}=\max_k\left(e^{\kappa_k}\frac{p_k}{p_t}\right)\in\mathbb Q_{>0}.
\]
Thus all interval endpoints are positive algebraic numbers, and $y=-\log s$ gives
\[
 E=\sum_j\theta_j\log\alpha_j,\qquad
 \theta_j\in\mathbb Q,\quad\alpha_j\in\overline{\mathbb Q}_{>0}.
\]
Choose a positive integer $M$ that clears the denominators of all $\theta_j$.
Then $P=\prod_j\alpha_j^{M\theta_j}$ is a positive real algebraic number, and
$P>1$, $P=1$, and $P<1$ correspond to $E>0$, $E=0$, and $E<0$, respectively.
Exact algebraic arithmetic and order comparisons complete the test.
\end{proof}

\paragraph{Thresholds for rational data.}
Under the assumptions of Proposition~\ref{prop:exacttest}, formula~\eqref{eq:thresholds} shows that finite thresholds
$\bar X_0,\bar Y_0$ have positive rational exponentials $e^{-\bar X_0},e^{-\bar Y_0}$.
They can therefore be used as the minimal base points.
When a threshold is $-\infty$ (the set where the envelope is negative is empty), supply a separate finite admissible base point with an algebraic exponential.

\section{Extension to multiplayer games}
\label{sec:multiplayer}
We extend the two-player comparison argument to $N$ players.
In place of a closed-form integral test, we define the capture region by whether the flow of a cooperative comparison system enters the positive region.

Consider a finite $N$-player game ($N\ge2$, $|S_i|\ge2$) with multilinear payoffs $u_i$,
under replicator dynamics applied to each player's mixed strategy on the product space
$\mathcal D_N^\circ=\prod_i\Delta(S_i)^\circ$.
Fix a strict pure Nash equilibrium $s^*$ and nonempty sets
$D_i\subseteq S_i\setminus\{s_i^*\}$, and write $N_i=S_i\setminus D_i$.
Assume that $s_i^*$ strictly outperforms each $k\in D_i$ against every pure profile in
$\prod_{j\ne i}N_j$, and that iterated elimination by strictly dominating pure strategies reduces $\prod_iN_i$ to $s^*$.
These conditions always hold when all alternative actions are included in the selected sets.

Define the ratio coordinates and constrained opponent polytopes by
\[
X_i(x)=\min_{k\in D_i}\log\frac{x_{i,s_i^*}}{x_{i,k}},\qquad
C_j(y_j)=\{z_j\in\Delta(S_j):z_{j,l}\le e^{-y_j}z_{j,s_j^*}
\ (l\in D_j)\}.
\]
The cooperative lower envelopes are
\[
F_i(y)=\min_{k\in D_i}\min_{z_j\in C_j(y_j),\ j\ne i}
\bigl[u_i(s_i^*,z_{-i})-u_i(k,z_{-i})\bigr].
\]
Each minimum can be taken over tuples of vertices.
The vertices of $C_j(y_j)$ are $e_b$ for $b\in N_j\setminus\{s_j^*\}$ and
\[
\frac{e_{s_j^*}+e^{-y_j}\sum_{l\in T}e_l}{1+|T|e^{-y_j}},
\qquad T\subseteq D_j.
\]
Thus $F_i$ is locally Lipschitz, independent of $y_i$, and nondecreasing in every other coordinate.
Because $F_i$ is bounded by the game's payoff differences,
$\dot y=F(y)$ has a unique global solution, denoted by $\Psi_\tau(y)$.
Set
\[
P^+=\{y:F_i(y)>0\text{ for all }i\},\quad
U=\{y:\Psi_{\tau_0}(y)\in P^+\text{ for some }\tau_0\ge0\},\quad
S_N=\{x:X(x)\in U\}.
\]

\begin{theorem}\label{thm:multi}
The set $S_N$ is nonempty, open, and forward invariant in the original product of simplex interiors.
Every trajectory starting in $S_N$ converges to $s^*$.
\end{theorem}
\begin{proof}
Multilinearity allows successive minimization over the opponents' vertices.
Their finitely many smooth parametrizations imply local Lipschitz regularity, and the nested polytopes imply monotonicity.
Along an actual trajectory, the ratio identity and differentiation of finite minima give
$D_+X_i\ge F_i(X)$.
The same comparison argument as in Remark~\ref{rem:order}
\cite[Theorem~3.2]{HirschSmith2006} therefore yields $X(\tau)\ge\Psi_\tau(X(0))$.
Within $P^+$, every coordinate of the comparison system increases, so its envelope values cannot decrease.
After entry at time $\tau_0$, every coordinate increases at a rate of at least
$\min_i F_i(\Psi_{\tau_0}(y))>0$.
The probabilities of dangerous actions therefore decay exponentially.
At each subsequent strict-dominance elimination step, the positive payoff gap on the surviving opponent profiles is perturbed by the exponentially vanishing mass on previously deleted actions.
The corresponding log ratio eventually increases at a positive rate.
Induction eliminates every action other than $s_i^*$.

As all opponent coordinates tend to infinity, the envelope limit is positive by the assumed payoff inequalities.
Hence $P^+$ contains a sufficiently large upper orthant, and $S_N$ contains the points of a relative neighbourhood of the target that lie in the original positive domain.
Continuous dependence and continuity of $X$ imply openness.
Order preservation makes $U$ an upper set; comparison and forward invariance of $U$ then imply forward invariance of $S_N$.
\end{proof}

\begin{proposition}[Explicit multiplayer boxes]\label{prop:multibox}
For a vector of box lower bounds $\theta\in\mathbb R^N$, write $r^-=\max\{-r,0\}$ and set
\[
\begin{aligned}
m_{ik}&=\min_{s_{-i}\in\prod_{j\ne i}N_j}
[u_i(s_i^*,s_{-i})-u_i(k,s_{-i})]>0,\\
M_{ik}&=\max_{s_{-i}\in\prod_{j\ne i}S_j}[u_i(s_i^*,s_{-i})-u_i(k,s_{-i})]^-,\qquad
\varepsilon_j=\frac{|D_j|e^{-\theta_j}}{1+|D_j|e^{-\theta_j}}
\end{aligned}.
\]
If every $\gamma_i:=\min_{k\in D_i}
\{m_{ik}-(m_{ik}+M_{ik})\sum_{j\ne i}\varepsilon_j\}$ is positive, then
$\{x:X_i(x)\ge\theta_i\ \forall i\}\subseteq S_N$.
Its normalized uniform product measure is
\[
\prod_i\sum_{T\subseteq D_i}\frac{(-1)^{|T|}}{1+|T|e^{-\theta_i}}.
\]
Moreover, if the selected sets include all alternative actions, every initial point in this box satisfies
$1-x_{i,s_i^*}(\tau)\le |D_i|e^{-\theta_i-\gamma_i\tau}$.
\end{proposition}
\begin{proof}
If $X_j\ge\theta_j$, opponent $j$ assigns at most $\varepsilon_j$ probability to dangerous actions.
By the union bound, the probability that at least one opponent chooses a dangerous action is at most $\sum_{j\ne i}\varepsilon_j$.
The payoff gap is at least $m_{ik}$ outside this event and at least $-M_{ik}$ on it.
Hence $F_i\ge\gamma_i>0$ throughout the box, which is mapped into $P^+$.
The measure follows by normalizing independent exponential random variables, as in Theorem~\ref{thm:volume}.
The time bound follows as in Theorem~\ref{thm:time}.
\end{proof}

\paragraph{Differences from the two-player case.}
Having $N$ comparison coordinates does not make vertex evaluation independent of the number of actions.
Entry into $P^+$ provides a certificate verifiable in finite time, but no finite stopping test is obtained for points outside $U$.
Numerical verification requires error control for both the comparison trajectory and $F$.

\section{How much of the basin does the capture region cover?}\label{sec:coverage}
This section examines how much of the target's basin is covered by the capture region. In two-action coordination games, the two sets coincide.
In general, the capture region is a subset of the basin, but every point in the basin enters the capture region in finite time.
Finally, numerical experiments on random games measure the gap between the two sets.

\paragraph{Two-action coordination games.}
\begin{proposition}[Equality in two-action coordination games]\label{prop:twobytwo}
Let $n=m=2$, and write $I=\{t,k\}$, $J=\{u,l\}$. Suppose that both $(t,u)$ and $(k,l)$ are strict
pure Nash equilibria. For any shifts, choosing the thresholds $X_0=\bar X_0$, $Y_0=\bar Y_0$ as the base points gives
\[
 \mathcal S=\{E>0\}=\operatorname{basin}(t,u),\qquad
 \{E<0\}=\operatorname{basin}(k,l).
\]
The set $\{E=0\}$ consists of the interior mixed equilibrium and the trajectories converging to it.
\end{proposition}
\begin{proof}
We have $D_r=\{k\}$, $D_c=\{l\}$, and $a=a_k>0$, $b=b_l>0$. Since $(k,l)$ is strict,
$c=c_{kl}<0$, $d=d_{lk}<0$.
The fraction in \eqref{eq:F} is strictly decreasing in $\rho$, so its infimum is attained at the upper endpoint $\rho=e^{-\mu-y}$, giving
\[
 F(y)=\frac{a+ce^{-\mu-y}}{1+e^{-\mu-y}},\qquad
 G(x)=\frac{b+de^{-\kappa-x}}{1+e^{-\kappa-x}}.
\]
Moreover, $q_l/q_u=e^{-\mu-Y}$, $p_k/p_t=e^{-\kappa-X}$, so the comparison inequalities become equalities:
$\dot X=F(Y)$, $\dot Y=G(X)$.
The functions $F,G$ are strictly increasing, with zeros $\bar Y_0=\log(-c/a)-\mu$, $\bar X_0=\log(-d/b)-\kappa$,
and $(\bar X_0,\bar Y_0)$ corresponds to the interior mixed equilibrium.

When $X>\bar X_0$, $Y>\bar Y_0$, both integrals are positive and $E>0$, so Theorem~\ref{thm:capture}
implies convergence to $(t,u)$. When $X\le\bar X_0$, $Y\le\bar Y_0$ and $(X,Y)\ne(\bar X_0,\bar Y_0)$,
we have $E<0$. Both $X,Y$ are nonincreasing and, after any positive time, decrease at rates bounded away from zero,
so the trajectory converges to $(k,l)$.
In the region $X<\bar X_0$, $Y>\bar Y_0$, we have $\dot E=|G(X)|F(Y)+|F(Y)|G(X)=0$,
while $X$ increases and $Y$ decreases. A trajectory that remains in this region is monotone and bounded, so it converges to a stationary point.
The only stationary point in the closure is $(\bar X_0,\bar Y_0)$, where $E=0$.
Thus, a trajectory with $E>0$ passes through $X=\bar X_0$, $Y>\bar Y_0$ in finite time and enters
$\{X>\bar X_0,\ Y>\bar Y_0\}$; a trajectory with $E<0$ passes through $Y=\bar Y_0$, $X<\bar X_0$ and enters
$\{X\le\bar X_0,\ Y\le\bar Y_0\}$; and a trajectory with $E=0$ converges to the mixed equilibrium.
The remaining region $\{X>\bar X_0,\ Y<\bar Y_0\}$ is treated in the same way by interchanging the players.
\end{proof}
This proposition recovers, from the present construction, the description of basin boundaries through a conserved quantity by Zhang and Hofbauer
\cite[Eqs.~(10)--(12)]{ZhangHofbauer2015} and the exact description of the Stag Hunt basin by Panageas and Piliouras
\cite[Theorem~5.1]{PanageasPiliouras2016}.
With three or more actions, the comparison generally gives inequalities, and the capture region is a proper subset of the basin.

\paragraph{Pullback.}
\begin{remark}[Relation to the full basin]\label{rem:pullback}
Any single capture region satisfies
$\operatorname{basin}(t,u)=\bigcup_{\tau\ge0}\phi_\tau^{-1}(\mathcal S)$.
Indeed, a convergent trajectory eventually enters a neighbourhood of the target, and $\mathcal S$ contains a relative neighbourhood of the target.
These preimages are nested, so their probabilities increase to the probability of the basin.
\end{remark}

\paragraph{Numerical setup.}
For each number of actions $n=2,\dots,6$, we generate 100 $n\times n$ bimatrix games with payoffs drawn independently
and uniformly from $[0,1]$, conditional on the existence of a strict pure Nash equilibrium, and select one such equilibrium uniformly as the target.
For each game, we draw 2000 initial states from the uniform prior. The selected sets include all nontarget actions, and the base points are the thresholds
$\bar X_0,\bar Y_0$ (when a threshold is $-\infty$, we use the union of the capture regions over finite base points).
The certificate integrals are evaluated numerically, branch by branch, using the primitives in Section~\ref{sec:computation}.
We compare the following four fractions (Table~\ref{tab:tightness}).
\begin{itemize}
\item \emph{Basin}: the fraction of initial states whose numerically integrated trajectories converge to the target. A trajectory is classified as converging to a strict pure equilibrium
as soon as it enters that equilibrium's invariant box (Theorem~\ref{thm:volume}); integration stops at time $1000$.
\item \emph{Box}: the fraction lying in the invariant box $\{X>\bar X_0,\ Y>\bar Y_0\}$ with zero shifts.
\item \emph{Capture region}: the fraction satisfying $E>0$ with zero shifts.
\item \emph{Shifted capture region}: the fraction satisfying $E>0$ for at least one of 16 choices of shifts for the dangerous actions.
To choose the shifts, let $h_l=\max_k c_{kl}^-/a_k$ measure how unfavorable the column action $l$ is to the row player's target, and set $e^{-\mu_l}=(h_l+0.05)^{-\gamma}$,
$\gamma\in\{0,\tfrac12,1,2\}$; the row shifts are chosen analogously. This choice constrains more unfavorable actions more tightly.
The certificate is valid for every choice of shifts, so the union is also valid.
\end{itemize}

\begin{table}[!htbp]
\centering
\begin{tabular}{crrrrr}
\hline
Actions $n$ & Basin & Box & Capture region & Shifted & Unresolved\\
\hline
2 & 92.3\% & 62.6\% & 92.3\% & 92.3\% & 0.0\%\\
3 & 83.2\% & 13.1\% & 47.5\% & 55.4\% & 8.0\%\\
4 & 68.1\% & 1.7\% & 7.2\% & 14.0\% & 15.3\%\\
5 & 55.9\% & 0.1\% & 0.6\% & 2.7\% & 28.2\%\\
6 & 56.9\% & 0.0\% & 0.1\% & 0.7\% & 30.2\%\\
\hline
\end{tabular}
\caption{Mean fractions across 100 random $n\times n$ games, with 2000 initial states per game.
``Unresolved'' denotes the fraction of initial states that neither entered the box of any strict pure equilibrium by the cutoff nor were classified as satisfying $E>0$.}
\label{tab:tightness}
\end{table}

\paragraph{Observations.}
\begin{itemize}
\item Among the $10^6$ initial states in total, none classified as satisfying $E>0$ converged to a different equilibrium under numerical integration.
Conversely, some initial states satisfied $E>0$ but remained unresolved by numerical integration at the cutoff
(122 for $n=2$ and 784 for $n=3$). These are counted in the basin column.
\item For $n=2$, the capture-region fraction equals the basin fraction, consistent with Proposition~\ref{prop:twobytwo}.
\item The effect of compensation is clear. The capture region covers approximately $3.6$ to $4.2$ times as many initial states as the box ($n=3,4,5$).
Shifts expand this further by a factor ranging from $1.2$ ($n=3$) to $4.4$ ($n=5$).
\item As the number of actions increases, however, the fraction covered by the capture region decreases rapidly. The ratio of the shifted capture-region fraction to the basin fraction
is approximately two-thirds for $n=3$, one-fifth for $n=4$, and one-twentieth for $n=5$.
\item The unresolved fraction increases with the number of actions (approximately 30\% for $n=6$).
These cases may correspond to cycling trajectories or trajectories with very slow convergence.
\end{itemize}

\paragraph{Entry times.}
By Remark~\ref{rem:pullback}, every trajectory converging to the target enters the capture region in finite time.
The same holds for any forward-invariant set containing a relative neighbourhood of the target.
Thus, when trajectories are integrated to certify convergence along the way, the value of the capture region lies in
how much earlier it is entered than a coarser set.
To measure this, we use 40 games for each $n=3,\dots,6$, with 1000 initial states per game,
and integrate to time $400$ using the fourth-order Runge--Kutta method with step size $0.05$.
For trajectories that enter the target's invariant box during this interval, we record the first entry time into each of the following four sets,
checking membership every $0.1$ time units:
the capture region with zero shifts, the union of the capture regions for the 16 shift choices above, the invariant box, and the rectangle
$\{p_t>\theta_r,\ q_u>\theta_c\}$, where $\theta_c=\max_k m_k/(a_k+m_k)$, $m_k=\max_l c_{kl}^-$,
and $\theta_r$ is defined analogously. Within this rectangle, the target action is a strict best response for each player,
regardless of the opponent's mixture.

\begin{table}[!htbp]
\centering
\small
\setlength{\tabcolsep}{4pt}
\begin{tabular}{crrrrrrrrr}
\hline
 & & \multicolumn{4}{c}{Fraction inside at time 0} & \multicolumn{4}{c}{Median entry time}\\
$n$ & Trajectories & Box & Rectangle & Capture region & Shifted & Box & Rectangle & Capture region & Shifted\\
\hline
3 & 30689 & 19.6\% & 16.1\% & 51.6\% & 62.5\% & 6.1 & 6.9 & 0.0 & 0.0\\
4 & 24618 & 1.6\% & 0.8\% & 9.5\% & 18.2\% & 17.1 & 17.5 & 12.1 & 6.0\\
5 & 21101 & 0.3\% & 0.1\% & 1.0\% & 4.2\% & 30.6 & 29.2 & 21.4 & 11.5\\
6 & 22743 & 0.2\% & 0.0\% & 0.5\% & 2.6\% & 49.0 & 46.0 & 34.8 & 19.8\\
\hline
\end{tabular}
\caption{First entry times into each set for trajectories that entered the target's invariant box by time $400$.
``Rectangle'' denotes the set defined by the target-probability bounds above; ``Shifted'' denotes the union of the capture regions for 16 choices of shifts.}
\label{tab:entry}
\end{table}

As Table~\ref{tab:entry} shows, for each of $n=4,5,6$, the median entry time into the unshifted capture region is approximately 70\% of the median box-entry time.
For the union over the 16 shift choices, it is approximately 40\%. Entry times into the rectangle are close to those into the box.
Although the fraction of initial states already in the capture region decreases rapidly with the number of actions, its advantage in certifying convergence earlier along trajectories
persists to a similar degree as the number of actions increases.
To turn certification based on numerically integrated trajectories into a rigorous proof, integration with certified error bounds is required.

\section{Limitations}\label{sec:summary}
\begin{itemize}
\item The certificate gives a sufficient condition. The fraction of the basin covered by the capture region decreases rapidly with the number of actions
(Section~\ref{sec:coverage}).
\item As the example in Section~\ref{sec:example} shows, convergence to the target is not global in general.
\item Exact membership decisions are limited to rational data and base points with algebraic exponentials. Section~\ref{sec:computation} counts
arithmetic operations; it does not give a polynomial upper bound on bit complexity.
\item The time bound (Theorem~\ref{thm:time}) applies when the selected sets include all nontarget actions.
\item In discrete time, we treat exponential-weights learning with exact payoff observations (Appendix~\ref{sec:hedge}).
The noisy case is discussed only in Remark~\ref{rem:hedge-noise}; learning in which players observe only the payoff of their own sampled action is not treated.
\item The size of the set of initial states converging to mixed equilibria is not addressed.
\end{itemize}

\appendix
\section{Discrete-time exponential-weights learning}\label{sec:hedge}
This appendix shows that the certificate extends to discrete-time exponential-weights learning (Hedge
\cite{FreundSchapire1997}) under explicit step-size conditions, without taking a continuous-time limit.
For the continuous-time limit yielding replicator dynamics, see \cite{MertikopoulosSandholm2016}.
Here, $n,m$ denote iteration indices. The noisy case is discussed in Remark~\ref{rem:hedge-noise}.

Fix a step size $\epsilon>0$ and consider the simultaneous updates
\[
 p_i(n+1)=\frac{p_i(n)e^{\epsilon(Aq(n))_i}}{\sum_{i'}p_{i'}(n)e^{\epsilon(Aq(n))_{i'}}},\qquad
 q_j(n+1)=\frac{q_j(n)e^{\epsilon(B^\top p(n))_j}}{\sum_{j'}q_{j'}(n)e^{\epsilon(B^\top p(n))_{j'}}}.
\]
Fix an admissible pair of action sets, shifts, and finite base points $X_0,Y_0$ as in Section~\ref{sec:receiver},
and denote the coordinate and certificate values at iteration $n$ by $X(n),Y(n),E(n)$.
Since $F,G$ are minima of finitely many branches with bounded derivatives, they are bounded and globally Lipschitz.
Let $\bar M$ be a common upper bound on $|F|,|G|$, and let $L$ be a common Lipschitz constant.

\begin{proposition}[Convergence under exponential-weights learning]\label{prop:hedge}
Suppose that $E(0)=e_0>0$, and set
\[
 y_*=\min\{y:\Lambda(y)\ge e_0/2\},\quad
 x_*=\min\{x:\Gamma(x)\ge e_0/2\},\quad
 \delta=\min\{F(y_*),G(x_*)\}>0,
\]
\[
 d_0=\max\{X_0-X(0),\ Y_0-Y(0),\ 0\}.
\]
If the step size satisfies
\[
 \epsilon\bar M\le\min\{x_*-X_0,\ y_*-Y_0\},\qquad
 L\bar M^2\,\epsilon\Bigl(\frac{d_0}{\delta}+\epsilon\Bigr)\le\frac{e_0}{2},
\]
then $E(n)\ge e_0/2$ for every $n$, and
$p_t(n)\to1$, $q_u(n)\to1$.
\end{proposition}
\begin{proof}
The update rule gives the exact log-ratio increment
$\xi_k(n+1)-\xi_k(n)=\epsilon\bigl((Aq(n))_t-(Aq(n))_k\bigr)$.
The right-hand side is $\epsilon$ times the fraction in \eqref{eq:F} and, as in the proof of Theorem~\ref{thm:capture},
is at least $\epsilon F(Y(n))$. The same argument for the column player gives
\begin{equation}\label{eq:discrete-comparison}
 X(n+1)\ge X(n)+\epsilon F(Y(n)),\qquad Y(n+1)\ge Y(n)+\epsilon G(X(n)).
\end{equation}
Since $\Gamma,\Lambda$ are nondecreasing,
$E(n+1)\ge\Gamma(X(n)+\epsilon F(Y(n)))+\Lambda(Y(n)+\epsilon G(X(n)))$.

Suppose that $E(m)\ge e_0/2$ for $m\le n$. At iteration $m$, there are three cases.

(i) $X(m)\ge X_0$ and $Y(m)\ge Y_0$. Since $F(Y(m))\ge0$, $G(X(m))\ge0$,
\eqref{eq:discrete-comparison} implies that neither coordinate decreases, and $E(m+1)\ge E(m)$.
Once this case is reached, it persists.

(ii) $X(m)<X_0$. Since $\Gamma(X(m))\le0$, we have $\Lambda(Y(m))\ge e_0/2$,
and hence $Y(m)\ge y_*$ and $F(Y(m))\ge\delta$.
The Lipschitz continuity of $|G|$ and $|F|$ gives
\[
 \Gamma(X(m)+\epsilon F)-\Gamma(X(m))\ge\epsilon F|G|-\tfrac12L\epsilon^2F^2,\qquad
 \Lambda(Y(m)+\epsilon G)-\Lambda(Y(m))\ge-\epsilon|G|F-\tfrac12L\epsilon^2G^2,
\]
where $F=F(Y(m))$, $G=G(X(m))$. The second inequality is for $G<0$; if $G\ge0$, its left-hand side is nonnegative.
Thus $E(m+1)\ge E(m)-L\bar M^2\epsilon^2$.
Also, $X(m+1)\ge X(m)+\epsilon\delta$, and the first step-size condition gives
$Y(m+1)\ge y_*-\epsilon\bar M\ge Y_0$. Thus the next iteration is in case (i) or (ii).

(iii) $Y(m)<Y_0$. Interchanging the players gives the same argument as in (ii), and the next iteration is in case (i) or (iii).

Cases (ii) and (iii) cannot occur simultaneously, since that would imply $E(m)\le0$, so the three cases are exhaustive.
Case (i) persists once reached; case (ii) cannot be followed by case (iii), nor case (iii) by case (ii).
Thus iterations outside case (i) form an initial consecutive block, all of type (ii) or all of type (iii).
In case (ii), $X$ increases by at least $\epsilon\delta$ at each iteration, so the number of such iterations is at most $d_0/(\epsilon\delta)+1$;
the same holds in case (iii). The total decrease in $E$ is therefore at most
$L\bar M^2\epsilon(d_0/\delta+\epsilon)\le e_0/2$, giving $E(n+1)\ge e_0/2$.
Induction shows that $E(n)\ge e_0/2$ for every $n$, and case (i) is reached after finitely many iterations.

After case (i) is reached, neither coordinate decreases, and at least one of the integrals is at least $e_0/4$.
For example, if $\Lambda(Y(m_1))\ge e_0/4$, then $F(Y(m))\ge F(Y(m_1))>0$ thereafter,
so $X(m)\to\infty$. Eventually $G(X(m))\ge G(+\infty)/2>0$, implying $Y(m)\to\infty$ as well.
The probabilities of dangerous actions decay geometrically. The remaining actions are eliminated by the same induction as in the proof of
Theorem~\ref{thm:capture}, using the exact log-ratio increment identity.
\end{proof}
The step-size conditions are always satisfied as $\epsilon\to0$. Thus exponential-weights learning converges to the target
from every point in the capture region when the step size is sufficiently small. The conditions can be imposed uniformly on compact subsets of $\mathcal S$.
Indeed, continuity of $E$ gives a positive lower bound on $e_0$ there; $x_*,y_*$ are nondecreasing in $e_0$,
and $F,G$ are nondecreasing, so $x_*-X_0$, $y_*-Y_0$, $\delta$ have positive lower bounds, while $d_0$ is bounded.
For the initial state in the example of Section~\ref{sec:example} ($e_0\approx1.18$), the proposition's conditions require approximately $\epsilon\le0.06$.
In numerical calculations, no decrease in $E$ was observed for $\epsilon\le1/2$; for $\epsilon=1,2$, the decreases
were only about $0.08$, $0.27$, respectively.

\begin{remark}[The noisy case]\label{rem:hedge-noise}
Suppose that the observation noise has conditional mean zero given the two players' shared history before the update,
and that the absolute noise difference between the target and each competing action is uniformly bounded by a constant $c>0$.
When the selected sets include all nontarget actions, the same argument extends to the stochastic setting.
One example is when each player updates using the payoff vector against the opponent's actually sampled action.
Fix a sufficiently small $r>0$, and consider the event that, over finitely many iterations, the absolute log-ratio fluctuations due to accumulated noise remain at most $r$.
Subtract the cumulative noise contribution from each shifted log ratio, subtract a further $r$, and take the minimum for each player.
The resulting coordinates lie below the actual $X,Y$ and satisfy \eqref{eq:discrete-comparison} on this event.
The probability of the complementary event is bounded by the Azuma--Hoeffding inequality.
After entry into a box near the target, each log ratio is a random walk with positive drift;
an exponential supermartingale and Ville's maximal inequality bound the probability of leaving the box.
Consequently, for a fixed initial state satisfying $E(0)>0$ and a sufficiently small step size $\epsilon$, the probability of failing to converge to the target
is at most $C\exp(-c'/\epsilon)$. The constants $C,c'>0$ depend on the game, the choice of capture region, the initial state, and $c$.
In 4000 numerical runs at each step size, using the learning rule in the example above and the initial state in Section~\ref{sec:example},
the fraction reaching a neighbourhood of the target by time $\epsilon n=60$ was $94.5\%$ for $\epsilon=1$, $98.3\%$ for $\epsilon=0.5$,
and $99.9\%$ for $\epsilon=0.2$; every run reached it for $\epsilon\le0.1$.
\end{remark}

\end{document}